\documentclass[
    ,final            
  ]
  {aipproc}

\layoutstyle{8x11single}

\begin{document}

\title{Search for Very High Energy Emission from Gamma-Ray Bursts using Milagro}


\author{P.~M.~Saz Parkinson for the Milagro Collaboration\footnote{
A.~Abdo,
B.~T.~Allen,
D.~Berley,
E.~Blaufuss,
S.~Casanova,
B.~L.~Dingus,
R.~W.~Ellsworth,
M.~M.~Gonzalez,
J.~A.~Goodman,
E.~Hays,
C.~M.~Hoffman,
B.~E.~Kolterman,
C.~P.~Lansdell,
J.~T.~Linnemann,
J.~E.~McEnery,
A.~I.~Mincer,
P.~Nemethy,
D.~Noyes,
J.~M.~Ryan,
F.~W.~Samuelson,
P.~M.~Saz~Parkinson,
A.~Shoup,
G.~Sinnis,
A.~J.~Smith,
G.~W.~Sullivan,
V.~Vasileiou,
G.~P.~Walker,
D.~A.~Williams,
X.~W.~Xu
and 
G.~B.~Yodh}}{
  address={Santa Cruz Institute for Particle Physics, University of California, Santa Cruz, CA 95064}
}

\begin{abstract}
Gamma-Ray Bursts (GRBs) have been detected at GeV energies by EGRET and 
models predict emission at > 100 GeV. Milagro is a wide field (2 sr) high duty cycle (> 90\%) ground 
based water Cherenkov detector that records extensive air showers in the 
energy range 100 GeV to 100 TeV. We have searched for very high energy emission 
from a sample of 106 gamma-ray bursts (GRB) detected since the beginning 
of 2000 by BATSE, BeppoSax, HETE-2, INTEGRAL, Swift or the IPN. No evidence for 
emission from any of the bursts has been found and we present upper limits from these bursts.
\end{abstract}

\maketitle


Some of the most important contributions to understanding gamma-ray bursts (GRBs) have 
come from observations of afterglows over a wide spectral range~\citep{2000ARA&A..38..379V}. 
Many GRB models predict very high energy (VHE, > 100 GeV) emission from GRBs at a level 
comparable to that at MeV energies (e.g.~\citep{dermer00,zhang01}). EGRET 
detected several GRBs at energies above 100 MeV, indicating that the spectrum of GRBs extends 
at least out to 1 GeV, with no evidence for a spectral cut-off~\citep{dingus01}. A second 
component 
was also found in one burst which extended up to at least 200 MeV and had a much slower temporal 
decay than the main burst~\citep{gonzalez03}. At very high energies, there has been no conclusive 
emission detected for any single GRB, though a search for counterparts to 54 BATSE bursts with 
Milagrito, a prototype of Milagro, found evidence for emission from one burst, with an after trials 
significance slightly greater than 3$\sigma$~\citep{atkins00a}. At these high energies, gamma 
rays are attenuated by the redshift-dependent extra-galactic background light 
(EBL)~\citep{primack05}, making the detection of VHE emission from GRBs very challenging. 

A search for an excess of events above those due to the background was carried out for each of the 
106 satellite-detected GRBs in our sample (see Table~\ref{grb_table}). These 
represent all the GRBs known to have occurred within the field of view of Milagro during its first 
seven years of operations (2000-2006)\footnote{GRB 060218, due to its long duration of more than 2000 s 
moved out of Milagro's field of view after the start of the burst. The limit presented here is for 
the initial 10 s hard spike reported by the instrument team.}. Milagro detected no significant emission from any of these bursts, and fluence 
upper limits are given in Table~\ref{grb_table}.\\



\begin{table}
\begin{tabular}{llllll}
\hline
GRB
&
Dur.& 
$\theta$
& 
z & 
Inst. & 
99\% UL \\
\hline
000113  & 370 	& 21	& ...	& BATSE 	& 5.5e-6 \\
000131$^1$ & 12 & 41 	& ...	& IPN 	 	& 6.5e-7 \\
000205  & 23 	& 25 	& ...	& BSAX  	& 6.9e-7 \\
000206	& 10 	& 39 	& ...	& BSAX 		& 9.3e-7 \\
000212  & 8     & 2.2	& ...	& BATSE 	& 1.1e-6 \\
000220	& 2.4	& 49 	& ...	& BATSE 	& 1.1e-5 \\
000226 	& 10    & 32	& ... 	& BATSE 	& 3.4e-6 \\
000226b$^1$& 94.5	& 32 	& ...	& IPN 	& 7.8e-7 \\
000301C	& 14    & 38	& 2.03 	& BATSE 	& ... \\
000302 	& 120   & 32	& ... 	& BATSE 	& 6.8e-6 \\
000314	& 12.8	& 45 	& ...	& BSAX 		& 3.6e-5 \\
000317 	& 550   & 6.4	& ... 	& BATSE 	& 7.9e-6 \\
000330 	&$0.2^*$& 30	& ... 	& BATSE 	& 1.0e-6 \\
000331 	& 55    & 38	& ... 	& BATSE 	& 1.2e-5 \\
000402 	& 120 	& 48 	& ...	& BSAX 		& 4.5e-5 \\
000408 	& 2.5   & 31	& ... 	& BATSE 	& 1.0e-6 \\
000424  & 5  	& 36 	& ...	& BATSE 	& 7.6e-7 \\
000508 	& 30    & 34	& ... 	& BATSE 	& 3.7e-6 \\
000607$^1$& 0.12& 42 	& ...	& IPN 		& 4.6e-7 \\
000615 	& 10    & 39	& ... 	& BSAX 		& 1.6e-6 \\
000630 	& 20    & 32	& ... 	& IPN	 	& 2.2e-6 \\
000707$^2$& 18 	& 43 	& ...	& IPN 		& 1.9e-6 \\
000707$^2$& 18 	& 41 	& ...	& IPN 		& 1.0e-6 \\
000727 	& 10    & 41	& ... 	& IPN	 	& 2.6e-6 \\
000730 	& 7     & 19	& ... 	& IPN	 	& 4.2e-7 \\
000821$^1$ & 8  	& 27 	& ...	& IPN 	& 6.9e-7 \\
000830$^1$ & 8  	& 35 	& ...	& IPN 	& 9.1e-7 \\
000926 	& 25    & 16	& 2.04  & IPN	 	& ... \\
001017 	& 10    & 42	& ... 	& IPN	 	& 2.2e-6 \\
001018 	& 31    & 32	& ... 	& IPN	 	& 2.1e-6 \\
001019 	& 10    & 20	& ... 	& IPN	 	& 1.1e-6 \\
001105 	& 30    & 8.5	& ... 	& IPN	 	& 1.4e-6 \\
001204  & 0.44	& 48 	& ...	& BSAX 		& 1.2e-5 \\
010104 	& 2     & 20	& ... 	& IPN	 	& 4.0e-7 \\
010220 	& 150   & 27	& ... 	& BSAX 		& 2.1e-6 \\
010613 	& 152   & 25	& ... 	& IPN 		& 2.9e-6 \\
010706	& 48 	& 37 	& ...	& IPN 		& 2.6e-6 \\
010903	& 41 	& 49 	& ...	& IPN 		& 2.9e-5 \\
010921 	& 24.6  & 10	& 0.45 	& HETE  	& {\bf 2.9e-5}$^\#$ \\
011130 	& 83.2  & 34	& ... 	& HETE 		& 3.4e-6 \\
011212 	& 84.4  & 33	& ... 	& HETE	 	& 6.7e-6 \\
020311  & 11.5 	& 27 	& ...	& IPN 		& 1.7e-7 \\
020429$^2$& 16	& 39 	& ...	& IPN 		& 4.6e-7 \\
020429$^2$& 16 	& 30 	& ...	& IPN 		& 3.0e-7 \\
020625b & 125 	& 38 	& ... 	& HETE  	& 5.7e-6 \\
020702  & 26 	& 34 	& ...	& IPN 		& 1.4e-6 \\
020908$^1$ & 17	& 19 	& ...	& IPN 		& 7.3e-7 \\
020914 	& 9	&  5.7 	& ...	& IPN 		& 4.2e-7 \\
021104 	& 19.7 	& 13  & ...  	& HETE 		& 7.5e-7\\
021112 	& 7.1 	& 34 	& ... 	& HETE 		& 9.4e-7 \\
021113 	& 20 	& 18 	& ... 	& HETE 		& 6.4e-7\\
021211 	& 6 	& 35  & 1.01 	& HETE 		& ... \\
030413 	& 15 	& 27 	& ... 	& IPN  		& 1.0e-6 \\
030823 	& 56 	& 33 	& ... 	& HETE 		& 2.8e-6 \\
\hline
\end{tabular}
\quad
\begin{tabular}{llllll}
\hline
GRB &
Dur. & 
$\theta$ & 
z & 
Inst. & 
99\% UL \\
\hline

031026	& 0.24 	& 45 	& ...	& IPN 		& 1.1e-6 \\
031220 	& 23.7 	& 43  & ... 	& HETE 		& 4.0e-6 \\
040506	& 175	& 49 	& ...	& IPN 		& 6.0e-6 \\
040924 	& 0.6 	& 43    & 0.859 & HETE  	& \textbf{1.4e-3}$^\#$ \\
041211 	& 30.2  & 43    & ... 	& HETE 		& 4.8e-6 \\
041219a & 520  	& 27 	& ... 	& INTGR. 	& 5.8e-6 \\
050124	& 4  	& 23	& ...	& Swift		& 3.0e-7 \\
050213 	& 17  	& 23 	& ...	& IPN 		& 1.3e-6 \\
050319  & 15   	& 45    & 3.24 	& Swift	  	& ... \\
050402  & 8   	& 40    & ... 	& Swift	  	& 2.1e-6 \\
050412  & 26   	& 37    & ... 	& Swift	  	& 1.7e-6 \\
050502  & 20   	& 43    & 3.793 & INTGR. 	& ...  \\
050504  & 80   	& 28    & ... 	& INTGR.  	& 1.3e-6 \\
050505  & 60   	& 29    & 4.3 	& Swift		& ...  \\
050509b & 0.128 & 10    & 0.226?& Swift	 	& \textbf{1.1e-6}$^\#$ \\
050522  & 15   	& 23    & ... 	& INTGR.	& 5.1e-7  \\
050607  & 26.5 	& 29    & ... 	& Swift	 	& 8.9e-7  \\
050703 	& 26	& 26  	&...	& IPN 		& 1.2e-6  \\
050712  & 35   	& 39    & ... 	& Swift	   	& 2.5e-6  \\
050713b & 30   	& 44    & ... 	& Swift	   	& 4.0e-6  \\
050715  & 52   	& 37    & ... 	& Swift	   	& 1.7e-6  \\
050716  & 69   	& 30    & ... 	& Swift	   	& 1.6e-6  \\
050820  & 20   	& 22    & 2.612 & Swift	  	& ...  \\
051103  & 0.17 	& 50 	& 0.001?& IPN		& \textbf{4.2e-6}$^\#$ \\ 
051109  & 36   	& 9.7   & 2.346 & Swift	  	& ... \\
051111  & 20   	& 44  & 1.55 	& Swift		& ...  \\
051211b & 80	& 33	& ...	& INTGR.	& 2.6e-6	\\
051221	& 1.4	& 42	& 0.55	& Swift		& \textbf{9.8e-4}$^\#$ \\
051221b	& 61	& 26	& ... 	& Swift		& 1.8e-6 \\
060102	& 20	& 40	& ... 	& Swift		& 2.0e-6 \\
060109	& 10	& 22	& ... 	& Swift		& 4.1e-7 \\
060110	& 15	& 43	& ... 	& Swift		& 3.0e-6 \\
060111b	& 59	& 37	& ... 	& Swift		& 2.3e-6 \\
060114	& 100	& 41	& ... 	& INTGR.	& 5.1e-6 \\
060204b	& 134	& 31	& ... 	& Swift		& 2.7e-6 \\
060210	& 5	& 43	& 3.91 	& Swift		& ... \\
060218	& 10	& 44.6	& 0.03 	& Swift			& \textbf{3.8e-5}$^\#$ \\
060306	& 30	& 46	& ... 	& Swift		& 7.2e-6 \\
060312	& 30	& 44	& ... 	& Swift		& 3.3e-6 \\
060313	& 0.7	& 47	& ... 	& Swift		& 2.7e-6 \\
060403	& 25	& 28	& ... 	& Swift		& 1.0e-6 \\
060427b	& 0.22	& 16	& ... 	& IPN		& 2.1e-7 \\
060428b	& 58	& 27	& ... 	& Swift		& 1.1e-6 \\
060507	& 185	& 47	& ... 	& Swift		& 1.8e-5 \\
060510b	& 330	& 43	& 4.9 	& Swift		& ... \\
060515	& 52	& 42	& ... 	& Swift		& 9.6e-6 \\
060712	& 26	& 35	& ... 	& Swift		& 3.8e-6 \\
060814	& 146	& 23	& ... 	& Swift		& 2.5e-6 \\
060904A	& 80	& 14	& ... 	& Swift		& 2.4e-6 \\
060906	& 43.6	& 29	& 3.685 & Swift		& ... \\
061002	& 20	& 45	& ... 	& Swift		& 4.0e-6 \\
061126	& 191	& 28	& ... 	& Swift		& 4.3e-6 \\
061210	& 0.8	& 23	& 0.41? & Swift		& \textbf{6.1e-6}$^\#$ \\
061222a	& 115	& 30	& ... 	& Swift		& 5.6e-6 \\
\hline
\end{tabular}
\caption{GRBs in the Milagro field of view (2000-2006). Column 1 is the GRB name. A superscript 
refers to the number of IPN error regions in the Milagro field of 
view. A superscript of one implies only one of two error regions fell in the Milagro field of view, 
while a two implies that both did, and they are listed one after the other. 
Column 2 gives the duration of the burst (in seconds), column 3 the zenith angle (in degrees), 
column 4 the measured redshift, column 5 the satellite(s) detecting the GRB, and column 6 
gives the Milagro 99\% confidence upper limit on the 0.2--20 TeV fluence in erg cm$^{-2}$. Numbers 
in bold (also labelled with a $^\#$) take into account absorption by the EBL (using the Primack 05 
model) for a redshift given in column 4. Those with three dots imply the redshifts are so high that 
all the emission is expected to be absorbed.\label{grb_table}}
\end{table}




\footnotesize{We acknowledge Scott Delay and Michael Schneider for their dedicated efforts in the 
construction and maintenance of the Milagro experiment.  This work has been supported by the 
National Science Foundation (under grants 
PHY-0245234, 
-0302000, 
-0400424, 
-0504201, 
-0601080, 
and
ATM-0002744) 
the US Department of Energy (Office of High-Energy Physics and 
Office of Nuclear Physics), Los Alamos National Laboratory, the University of
California, and the Institute of Geophysics and Planetary Physics.}




\bibliographystyle{aipproc}   

\end{document}